\documentclass{aa}

\usepackage{graphicx}
\usepackage{color}
\usepackage{amssymb}
\usepackage{enumerate}

\newcommand{\HII}{H\,{\sc ii}}
\def\p0{\phantom{0}}

\begin{document}

\title{The $\Sigma-D$ relation for planetary nebulae}

\author{D. Uro{\v s}evi{\' c}\inst{1,5},
B. Vukoti{\' c}\inst{2}, B. Arbutina\inst{1},
D. Ili{\' c}\inst{1,5}, M. Filipovi{\' c}\inst{3},
I. Boji\v{c}i\'c\inst{4}, S. {\v S}egan\inst{1}, \and S. Vidojevi{\' c}\inst{1}}

\institute{Department of Astronomy, Faculty of Mathematics,
University of Belgrade, Studentski trg 16, P.O. 550, 11000
Belgrade, Serbia \and Astronomical Observatory,
Volgina 7, 11160 Belgrade 74, Serbia \and University of Western
Sydney, Locked Bag 1797, Penrith South DC, NSW 1797, Australia \and
Department of Physics, Macquarie University, Sydney, NSW 2109, Australia \and
Isaac Newton Institute of Chile, Yugoslavia Branch}

\offprints{Dejan Uro\v sevi\' c, \email{dejanu@matf.bg.ac.yu}}

\abstract{We present an extended analysis of the relation between
radio surface brightness and diameter -- the so-called $\Sigma-D$
relation for planetary nebulae (PNe). We revise our
previous derivation of the theoretical $\Sigma-D$ relation for
the evolution of bremsstrahlung surface brightness in order to
include the influence of the fast wind from the central star.
Different theoretical forms are derived: $\Sigma\propto D^{-1}$
for the first and second phases of evolution and $\Sigma\propto
D^{-3}$ for the final stage of evolution. Also, we analyzed
several different Galactic PN samples. All samples
are influenced by severe selection effects, but Malmquist
bias seems to be less influential here than in the supernova
remnant (SNR) samples. We derived empirical  $\Sigma-D$ relations for 27 sample sets
using 6 updated PN papers  from which an additional 21 new
sets were extracted.  Twenty four of these have a trivial form of
$\beta\approx2$.   However, we obtain one empirical $\Sigma-D$ relation that may be useful for determining
distances to PNe. This relation is obtained by extracting a recent nearby ($<$ 1 kpc) Galactic PN sample.

}

\date{}

\titlerunning{The $\Sigma-D$ relation for planetary nebulae}
\authorrunning{D. Uro\v sevi\'c et al.}

\maketitle

\section{Introduction}

Planetary nebulae (PNe) are usually identified by their optical
spectra, consisting mainly of recombination lines of hydrogen,
helium and  collisionally excited light elements.
They are composed of gaseous shells ionized
by a hot central star, which allow for the necessary
conditions to create  their spectra.
PNe were first discovered over 200 years ago. Today,
 there are more than
 { 2500}  confirmed Galactic PNe with total number
 estimates at least ten times higher.
PNe emission from radio to \mbox{X-ray} have
been detected over the past 30 years.
Several hundreds of these PNe have been observed in
radio-continuum alone. Their number far
outrank the known Galactic supernova
remnant (SNR) population (approximately 250).

The relation between radio surface brightness and diameter for
SNRs, the so-called $\Sigma-D$ relation, has been the subject of
extensive discussions over the past forty years. This $\Sigma-D$
relation has a power law form:

\begin{equation}
\Sigma\propto D^{-\beta},
\end{equation}

\noindent where $\Sigma$ is the radio surface brightness, $D$ is
diameter and $\beta$ the slope of the log-log plot.

Observational  improvements, including the
increased use of radio interferometers,
account for the high number of resolved radio PNe over the past two decades.
 The $\Sigma-D$ relation for PNe has
only been sporadically discussed (for example, \cite{Amnuel84}) until
\cite{Uro07} (hereafter Paper I).

The statistical analysis of the Galactic PN distances
was presented in \cite{Daub82} and in \cite{Cahn92three}.
Using radio data, the primary statistical method for determining distance
was based on the correlation between PN radius and brightness temperature,
known as the $R-T_{\rm b}$ relation\footnote{This relation is essentially equivalent to the
$D-\Sigma$ relation.}. \cite{Van95} derived
an empirical $R-T_{\rm b}$ relation for 131 Galactic PNe,
felt to be located near the Galactic center, at
approximately the same distance.
Additionally, a smaller sample of 23 (non-Galactic bulge) PNe were studied
having  well-determined distances. The same calibration
relations were obtained in both cases. \cite{Zhang95} studied the
statistical distance scale for PNe based on another sample of 132 PNe
having well-determined individual distances. He found that the
derived distance scale was in good agreement with those derived by
\cite{Van95}.
More recently,
\cite{Phillips02} (hereafter Ph02) derived
$R-T_{\rm b}$ relations for 44 nearby PNe whose distances are less than 0.7~kpc.
Thus,
different samples of Galactic PNe with known distances were
defined in the three previously cited papers. All of these
derived $R-T_{\rm b}$ relations were
applied to PNe with unknown distances.
{ We also note that a recent paper by \cite{Stanghellini08} (hereafter SSV)
gives an updated reliable sample of individual Galactic PN   distances.}



The statistical analysis of Galactic
radio PNe is less developed than
those of Galactic SNRs\footnote{The radio samples of Galactic PNe are richer in terms of the number of objects, but incompleteness  is drastically higher than in SNR samples --  less than 5\% of total number of PNe can be detected by radio surveys (\cite{Kwok94}).}. Therefore,   selection effects are felt to greatly influence Galactic radio PN samples causing the statistical
determination of distance  to be highly uncertain.

\smallskip

In the present paper, we use the methodology for the statistical analysis of SNRs to improve
that for Galactic radio PNe.
Primary objectives include the following:

\begin{enumerate}[(i)]

\item Development  of an improved form of the theoretical $\Sigma-D$ relation
for PNe (compared to the one derived in Paper I) by incorporating  the
interaction between asymptotic-giant-branch (AGB) star wind and
the fast wind from the central star.

\item Derivation of improved empirical  $\Sigma-D$ relations using
current and previous samples to determine if these
are all under the influence of selection effects.

\item Validation of $\Sigma-D$ relations for the
determination of radio PN distance.

\end{enumerate}


\section{Dynamics of PNe}

{ In considering the dynamics of PNe, we adopt the {\it interacting stellar
winds model }(ISW; \cite{Zhang93}, \cite{Kwok94}, and references therein).  This model implies that a nebular shell
results from the sweeping up of circumstellar material moving
within the AGB star wind by a fast wind emanating from the PN central star. The
speed of the fast wind is 100 times that of the AGB wind and acts as a
"snowplow", piling  up matter into a high density shell.
 This interaction produces
two shocks.  The inner shock is formed  at the interaction between central star
wind and shell.  The outer one is situated in the AGB wind just outside the shell.
  Most of the volume interior of the shell is made up
of shocked central-star wind at a high temperature (millions of
degrees). The outer shock is likely to be isothermal -- similar to the
shocks in the radiative phase of SNR evolution,   where a dense
shell is formed.

The dependence between gas density and
radius of a PN is necessary for deriving the theoretical
$\Sigma-D$ relation for PNe. This dependence was shown in Paper~I
for a model consisting of one steady wind emanating from the central star. A  trend
of decreasing radio surface brightness with increasing diameter
 was shown in Paper~I. The radiation mechanism used for
derivation of the theoretical $\Sigma-D$ relation is thermal
bremsstrahlung, which is the basic process of
radio-continuum emission in \HII\ regions.}

We begin our derivation using the following notation:\\

$R_s$ -- shell radius,\\

$R_o$ -- outer radius of the shell,\\

$R_i$ -- inner radius of the shell,\\

$V_s$ -- velocity of the shell,\\

$V$ -- AGB wind velocity,\\

$v$ -- fast wind velocity,\\

$\dot{M}$ -- mass loss in the AGB phase,\\

$\dot{m}$ -- fast wind mass loss,\\

$M_s$ -- mass in the shell,\\

$\rho_s$ -- density of the shell,\\

$t=0$ -- the moment when the AGB wind stops,\\

$t=\tau$ -- the moment when the fast wind starts.\\

\subsection{The momentum conserving phase}

If we assume momentum conservation during the interaction, then:

\begin{equation}
P_s= M_s V_s = P_{\scriptscriptstyle AGB}+P_{\scriptscriptstyle
FW},
\end{equation}

\begin{eqnarray}
P_s &=& P_{\scriptscriptstyle AGB}+P_{\scriptscriptstyle FW}
\nonumber \\
&=& \int ^{R_o} _{Vt} 4 \pi r^2 \rho _{\scriptscriptstyle AGB} V
\mathrm{d}r + \int _{R_i} ^{v(t-\tau)} 4 \pi r^2 \rho
_{\scriptscriptstyle FW} v \mathrm{d}r \nonumber \\
&=& \int ^{R_o} _{Vt} 4 \pi r^2 \frac{\dot{M}}{4 \pi r^2 V} V
\mathrm{d}r + \int _{R_i} ^{v(t-\tau)} 4 \pi r^2 \frac{\dot{m}}{4
\pi r^2 v} v \mathrm{d}r \nonumber \\
&=&  {\dot{M}} \big(R_o - Vt\big) + {\dot{m}}\big(v(t-\tau) -
R_i\big),
\end{eqnarray}

\noindent while the mass in the shell is:

\begin{eqnarray}
M_s &=& M_{\scriptscriptstyle AGB}+M_{\scriptscriptstyle FW}
\nonumber \\
&=& \int ^{R_o} _{Vt} 4 \pi r^2 \rho _{\scriptscriptstyle AGB}
\mathrm{d}r + \int _{R_i} ^{v(t-\tau)} 4 \pi r^2 \rho
_{\scriptscriptstyle FW} \mathrm{d}r \nonumber \\
&=& \int ^{R_o} _{Vt} 4 \pi r^2 \frac{\dot{M}}{4 \pi r^2 V}
\mathrm{d}r + \int _{R_i} ^{v(t-\tau)} 4 \pi r^2 \frac{\dot{m}}
{4\pi r^2 v}  \mathrm{d}r \nonumber \\
&=&  \frac{\dot{M}}{V} \big(R_o - Vt\big) +
\frac{\dot{m}}{v}\big(v(t-\tau) - R_i\big).
\end{eqnarray}

\noindent If, additionally, the shell is thin: $R_i \rightarrow
R_o = R_s$ then we may write:

\begin{equation}
P_s =  \big( {\dot{M}} - {\dot{m}} \big) R_s - \dot{M}Vt +
\dot{m}v(t-\tau),
\end{equation}

\begin{equation}
M_s =  \big( \frac{\dot{M}}{V} - \frac{\dot{m}}{v} \big) R_s -
\dot{M}t + \dot{m}(t-\tau)
\end{equation}

\noindent (see Equation 6 in \cite{Zhang93}).

\noindent If $V_s$ = const., the shell is expanding following the
law:

\begin{equation}
R_s =  \frac{vV\tau }{v-V} + V_s \big( t - \frac{v\tau}{v-V} \big),
\end{equation}

\noindent which may be reduced to Equation (5) in \cite{Zhang93p}  if $v \gg V$:

\begin{equation}
R_s \approx  {V\tau } + V_s \big( t - {\tau} \big).
\end{equation}

\noindent If $t \gg \tau$ then,

\begin{equation}
R_s \approx  V_s t ,
\end{equation}

\noindent and the shell velocity can be expressed as:

\begin{equation}
V_s = \frac{P_s}{M_s} = \frac{ \big( {\dot{M}} - {\dot{m}} \big)
R_s  + (\dot{m}v - \dot{M}V)\frac{R_s}{V_s} } {\big(
\frac{\dot{M}}{V} - \frac{\dot{m}}{v} \big) R_s  + (\dot{m} -
\dot{M})\frac{R_s}{V_s} }.
\end{equation}

\noindent Solving this equation we obtain:

\begin{equation}
V_s = \frac{ {\dot{M}} - {\dot{m}} + (v - V) \sqrt{\frac{\dot{M}
\dot{m}}{vV}} }{\frac{\dot{M}}{V} - \frac{\dot{m}}{v}},
\end{equation}

\noindent which is Equation  (4)  in \cite{Zhang93p}.

The average density inside the shell can be expressed as:

\begin{equation}
\rho_s = \frac{M_s }{\frac{4\pi}{3}f R_s^3},
\end{equation}

\noindent where $f$ is the volume filling factor. From Equations
(6) and (9) we see that if $t \gg \tau$:

\begin{equation}
M_s = \Big[ {\big( \frac{\dot{M}}{V} - \frac{\dot{m}}{v} \big)   +
\frac{\dot{m} - \dot{M}}{V_s}} \Big]R_s.
\end{equation}

\noindent If $\dot{M} \gg \dot{m}$, and $v \gg V$,
then $M_s \approx (\frac{\dot{M}}{V}+\frac{\dot{m} - \dot{M}}{V_s})R_s$ and

\begin{equation}
\rho_s \propto \frac{\dot{M}}{{4\pi}R_s^2 V}.
\end{equation}

\noindent Essentially, we have   perturbed the AGB wind ($\rho_{\scriptscriptstyle AGB} =
\frac{\dot{M}}{{4\pi}R_{\scriptscriptstyle AGB}^2 V}$). From Equation (11) we estimate $V_s
\approx 15\ \mathrm{km\ s}^{-1}$ using   typical values of
$\dot{M} = 10^{-5}\ M_{\odot}\mathrm{yr}^{-1},\ \dot{m} = 10^{-8}\
M_{\odot}\mathrm{yr}^{-1},\ V = 10\ \mathrm{km\ s}^{-1}  \mathrm{and} \ v =
2000\  \mathrm{km\ s}^{-1}$. We obtain the dependance ${\rho_s \propto R_s^{-1}}$
using Equations (12) and (13) assuming that the
shell thickness $\Delta = R_o - R_i$ = const. and $f = 1 -
(\frac{R_i}{R_o})^3 = 1 - (1-\frac{\Delta}{R_s})^3 \approx
3\Delta/R_s$.

\subsection{The energy conserving phase}

The momentum-conserving phase is perhaps, a reasonable description of
early PN evolution. However, if a part of the energy of the
fast wind is transformed into thermal energy, then the pressure of the hot
bubble  ($P$) will provide additional acceleration to the nebular
shell. In this ``energy-conserving'' phase, the equation of motion
and the energy equation is expressed as:
\begin{equation}
M_s \frac{\mathrm{d}^2 R_s}{\mathrm{d}t^2} = 4 \pi R_s^2 P -
\frac{\dot{M}}{V}(V_s - V)^2,
\end{equation}
\begin{equation}
\frac{\mathrm{d} }{\mathrm{d} t}\Big( \frac{1}{\gamma -1} P\cdot
\frac{4\pi}{3}R_s^3 \Big) = \frac{1}{2}\dot{m}v^2 - 4 \pi R_s^2 P
V_s
\end{equation}

\noindent (\cite{Kwok00}).

The similarity solution method can be used for defining the change
of PN radius with time. The power-law form $R_s\propto t^\alpha$, is expected
for a PN in the energy conserving phase of evolution. We can define
dimensionless variable in form:

\begin{equation}
\xi\equiv R_st^l\rho_{\scriptscriptstyle AGB}^m\dot{E}^n,
\end{equation}

\noindent where $\dot{E}$ is the power of fast wind which injects energy (through the inner shock) into the hot bubble. After similarity analysis, the following expression is obtained:

\begin{equation}
R_s=\xi\bigl({\dot{E}\over\rho_{\scriptscriptstyle AGB}}\bigr)^{1\over5}t^{3\over5}.
\end{equation}

\noindent We obtain a $R_s\propto t$ $(\alpha=1)$ dependence assuming $\rho_{\scriptscriptstyle AGB}\propto R_s^{-2}$ and using Equation (18).  Finally, a
stationary solution $V_s=\dot{R_s}$ = const. can be used for further derivations.


The mass of the compressed AGB wind in the shell, using $V_s$=const., is:
\begin{equation}
M_s = \dot{M}\Big( \frac{1}{V} -\frac{1}{V_s} \Big) R_s.
\end{equation}
\noindent If for $f$, we assume   $\Delta$ = const., then Equation (12) implies
$\rho_s \propto R_s^{-1}$.

\subsection{The final phase}

We only have   AGB wind momentum and mass conservation in the final phase of evolution when the pressure of the hot bubble, $P\rightarrow 0$:
\begin{eqnarray}
P_s &=& \int ^{R_s} _{R'} 4 \pi r^2 \rho _{\scriptscriptstyle AGB}
V \mathrm{d}r + P' \nonumber \\
&=&  {\dot{M}} \big(R_s - R'\big) + P',
\end{eqnarray}

\begin{eqnarray}
M_s &=&  \int ^{R_s} _{R'} 4 \pi r^2 \rho _{\scriptscriptstyle
AGB} \mathrm{d}r + M' \nonumber \\
&=&  \frac{\dot{M}}{V} \big(R_s - R') + M',
\end{eqnarray}
where $P'$ and $M'$ are the momentum and mass acquired by the time
the shell radius reaches some value $R'$. For the shell
velocity we now have:

\begin{equation} V_s = \frac{P_s}{M_s} = \frac{{\dot{M}} \big(R_s - R'\big) + P'}{{\dot{M}} \big(R_s - R'\big) + M'V}V\
\rightarrow \ V,
\end{equation}
as $R_s \rightarrow \infty$.

During the early phases, the PN shell tends to the contact
discontinuity, while in the last phase, it will be just behind the
isothermal forward shock. If we apply isothermal shock jump
conditions, we can estimate:

\begin{equation} \rho_s = \mathrm{M}^2 \rho _{\scriptscriptstyle AGB},
\end{equation}
where $\mathrm{M} = V_s/ c_s$ is the Mach number and $c_s$, the
isothermal sound speed. The density jump is not limited to a maximum
value of 4, as in an adiabatic shock.  If $V_s \approx $ const.,
then $\rho \propto R_s^{-2}$ behind the shock.

\section{Theoretical $\Sigma -D$ relation for PNe}

Assuming a thermal bremsstrahlung mechanism is responsible for
radiation of \HII\ regions at radio wavelengths, we may write the
volume emissivity $\varepsilon_\nu$ of a PN as:

\begin{equation}
\varepsilon_\nu [\mbox{ergs~s$^{-1}$~cm$^{-3}$~Hz$^{-1}$}] \propto
{n^2 T^{-1/2}},
\end{equation}

\noindent  (\cite{Rohlfs96}) where $n$ is the volume density and $T$   the electron
temperature.

The surface brightness can be expressed as:

\begin{equation}
\Sigma_\nu \propto \frac{L_{\nu }}{D^2} \propto \varepsilon_\nu D,
\end{equation}

\noindent where $L_{\nu}$ is luminosity and $D=2R_s$,   the
diameter of the PN. We combine Equations (24) and (25),
use the derived dependence from
Section 2  such that  $n \propto
\rho _s \propto D^{-1}$, and assume that
  temperature is constant.\footnote {\HII\  regions are approximately
isothermal at $T\sim10^4$ K. } This radio surface brightness to diameter relation
has a different form for the first and second phase of evolution (in comparison to the earlier result given in
Paper I):

\begin{equation}
\Sigma\propto n^2T^{-1/2} D\propto D^{-1}.
\end{equation}


\noindent Generally, we may write $n\propto D^{-x}$, where $x\ga 1$ (if
$\Delta$ = const.).

We do not expect the temperature to be
strictly constant throughout the nebula. There are temperature
gradients in PNe arising from radiation hardening. More energetic
photons will travel further and when they are absorbed by the PN,
they will impart a greater kinetic energy to the ions thereby
producing a higher temperature. This would  only slightly
increase the value for $\beta$ (+0.1 approximately), although the
$T_e - D$ dependence is not quite of the power-law form (see
  numerical model results from \cite{Evans85}).

The value
$\beta = 1$   under the above  conditions (including
$\Delta$ = const.) is only a theoretical lower limit; the
$\Sigma-D$ relation could be steeper due to the these effects, as
can be seen from Equation (26). The  relation given here is derived under the
assumption of spherical symmetric expansion. However, we know that
the large number of PNe show a high asymmetry and can have a very
complex morphology. This may explain why the theoretical $\Sigma-D$ relation
is not necessary connected to the empirical relations that we
discuss in Section~4. Even if we had a spherically symmetric PNe with
$\Sigma \propto D^{-1}$, the constant of proportionality would be
different for different phases of evolution, as can be inferred from Section~2.
Moreover, parameters governing the evolution ($V$, $v$, $\dot{M}$,
$\dot{m}$) are not the same for all PNe, so no unique empirical
relation for PNe is feasible.

In the final stage of evolution, after the fast
wind has ceased, the isothermal shock wave continuously perturbs AGB wind.
Nonetheless, the surface brightness steeply declines,  as demonstrated
in Paper I, because the mechanical energy input (acceleration)
from the central star no longer exists. Using
$\rho_{\scriptscriptstyle AGB}\propto R_s^{-2}$, the $\Sigma-D$
relation in the final stage of   PN evolution can be expressed as:

\begin{equation}
\Sigma\propto D^{-3}.
\end{equation}

\section{The empirical $\Sigma-D$ relation for PNe}

The most important prerequisite for defining a proper empirical
$\Sigma-D$ relation is the extraction of a valid sample of PNe.  This
sample should consist of calibrators having well determined distances.


The distances to the calibrators have to be determined by reliable
methods.  These may include trigonometric or spectroscopic parallaxes of PN central
stars and methods based on nebulae expansion.
Although limited by current technology, trigonometric parallax    is the most direct method of
measuring distances.

All extracted PN samples suffer from
severe selection effects  caused by limitations in survey
sensitivity and resolution. For Galactic samples, Malmquist
bias\footnote{Intrinsically bright PNe are favored in any flux limited survey
because they are sampled from a larger spatial volume.} may be the most severe
effect; this is similar to the situation for Galactic  SNR samples (\cite{Uro05}).





For the literature extracted PN samples in this paper,  we used   flux densities at 5 GHz.
This is because most PNe are expected to be optically thin at this frequency (see
Section 5). Radio diameters would be preferable, however for most PNe
used here, only optical diameters are available.



{ We extracted   27  PN sample sets as shown in
Tables~1 and 2. While both detail  fit parameters for $\Sigma-D$
and $L-D$, Table 1 does so by author (18 PN  sets) and  Table~2
by the specific distance determination  method used (9 PN   sets).

Data for  sample Nos.  1 --  4 in Table 1 are
taken from   SSV and \cite{Cahn92}. We used Equation (7) from \cite{Van95}
to convert the brightness temperatures listed in Ph02 to their 5 GHz flux.
PN distance ranges (kpc) are listed in Table 1 (e.g. { ``$1.0<d\leq2.0"$} indicates that
all PNe in that set have distances between 1 and 2~kpc).
Three PNe (He~1-5, HDW~6 and HaWe~5) indicated by ``\dag"  were removed
from some   sets because of inconsistencies with other data, as we will detail
in Section~5.

    Sample set 15 is taken from \cite{Zhang95} and consists of
132 PNe with well-determined individual distances. Set 16
contains 23 PNe with well-determined distances, while, set 17
consists of 132 Galactic bulge PNe. Both samples were taken from \cite{Van95}.
Set 14 consists of 109 Galactic
bulge PNe taken from \cite{Bensby01}.
Set 18 (88 PNe) is actually the \cite{Van95}
Galactic bulge sample with stronger criteria applied to extract Galactic bulge PNe as
 defined in \cite{Bensby01}.

 The first eight
sample sets listed in   Table 2  were obtained from SSV and
Ph02 data.  Set 9,   USNO-PN, consists of   PNe with
well-measured trigonometric parallaxes,   determined by
United States Naval Observatory (USNO) ground based observations
(\cite{Harris07}). Details of this USNO-PN sample data are
presented in Table~3.

The trigonometric parallaxes of PNe
determined by the Hipparcos mission suffer from large measurement
errors (\cite{Harris07}) and therefore are of limited practical
use for our study. The average measurement error in Hipparcos
trigonometric parallaxes for some of our PNe is 57 \%.
Thus, the USNO-PN sample set is more reliable and have more
accurate distance measurents\footnote{An average error of measured
trigonometric parallaxes in this sample is 17.2 \%.}.
The $\Sigma-D$ relation for the USNO-PN sample has the following form:}

\begin{equation}
\Sigma = 7.39^{+4.50}_{-2.79}10^{-22}D^{-2.38\pm0.56}.
\end{equation}

\noindent Corresponding $\Sigma_\nu-D$ and $L_\nu-D$ diagrams
are shown in Fig.~1.

\begin{table*}

\caption[]{The results of the $\Sigma-D$ and $L-D$ fits, where the
parameters of the fit ($\beta$ and $\alpha$, respectively) and the
correlation coefficient $r$ are given for each sample. The number
of PNe in the sample is given in the last column ($N$).
  SSV stands for \cite{Stanghellini08}, B\&L for  \cite{Bensby01} and Ph02 for \cite{Phillips02}. }

\smallskip

\centerline{\begin{tabular} {llccccc}
\hline \hline \noalign{\smallskip}
No.& Sample$^{a}$&
$\beta_{_{\Sigma - D}}$  &$r_{_{\Sigma- D}}$&
$\alpha_{_{L-D}}$ & $r_{_{L - D}}$ & $N$\\
\noalign{\smallskip} \hline \noalign{\smallskip}
01 & SSV  $d<$1.0 &2.61$\pm$0.21 & $-0.97$ &  \p00.61$\pm$0.21 & $-0.67$ & \p013\\
02 & SSV  1.0$\leq d<$2.0 &1.89$\pm$0.43 & $-0.78$ &  --0.11$\pm$0.43 & \p00.08 & \p015\\
03 & SSV  $d\geq$2.0 &3.96$\pm$0.62 & $-0.91$ &  \p01.97$\pm$0.61 & $-0.73$ & \p011\\
04 & SSV   &2.56$\pm$0.22 & $-0.89$ &  \p00.56$\pm$0.22 & $-0.40$ & \p039\\
05 & SSV + Ph02    $d\leq$0.4  &1.47$\pm$0.40 & $-0.74$ &  --0.53$\pm$0.40 & \p00.38 & \p013\\
\noalign{\smallskip}
06 & SSV + Ph02  $0.4<d\leq$0.6   &2.18$\pm$0.22 & $-0.91$ &  \p00.18$\pm$0.22 & $-0.17$ & \p023\\
07 & SSV + Ph02 0.6$<d\leq$1.0   &2.37$\pm$0.20 & $-0.95$ &  \p00.37$\pm$0.20 & $-0.43$ & \p018\\
08 & SSV + Ph02 1.0$<d\leq$2.0 & 2.64$\pm$0.40 & $-0.88$ &   \p00.64$\pm$0.40 & $-0.41$ & \p015\\
09 & SSV + Ph02 $d>$2.0 & 4.44$\pm$0.47 & $-0.96$ &   \p02.44$\pm$0.47 & $-0.90$ & \p0\p09 \\
10 & SSV + Ph02 &2.40$\pm$0.17 & $-0.86$ &  \p00.40$\pm$0.17 & $-0.27$ & \p078\\
\noalign{\smallskip}
11 & SSV + Ph02 $d\leq$0.4 \dag    & 2.31$\pm$0.38 & $-0.90$ &  \p00.31$\pm$0.38 & $-0.26$ & \p011\\
12 & SSV + Ph02 0.4$<d\leq$0.6 \dag   & 2.33$\pm$0.15 & $-0.96$ &  \p00.33$\pm$0.15 & $-0.44$ & \p022\\
13 & SSV + Ph02 \dag &2.57$\pm$0.14 & $-0.92$ &  \p00.57$\pm$0.14 & $-0.45$ & \p075\\
14 &  B\&L  &2.31$\pm$0.11 & $-0.90$ &  \p00.31$\pm$0.11 & $-0.26$ & 109\\
15 & \cite{Zhang95}       & 2.17$\pm$0.14 & $-0.80$ &  \p00.17$\pm$0.14 & $-0.10$ & 132\\
\noalign{\smallskip}
16 &\cite{Van95} & 2.41$\pm$0.34 & $-0.84$ &  \p00.41$\pm$0.34 & $-0.26$ & \p023\\
17 & \cite{Van95} &  2.24$\pm$0.09 & $-0.92$ & \p00.24$\pm$0.09 & $-0.24$ & 132 \\
18 & \cite{Van95} (stronger criteria by B\&L)&  2.21$\pm$0.09 & $-0.94$ & \p00.21$\pm$0.09 & $-0.25$ & \p088 \\
\noalign{\smallskip} \hline
\end{tabular}}

\smallskip
$^{a}${PNe from Ph02 are included in the sample sets (5 -- 13) as additional,
if they were not extracted by the SSV in their sample; $d<1$   denotes that
a set of PNe have distances less than 1 kpc, $1\leq d<$2, between 1 and 2~kpc, etc.\\
\dag\, --\, 3 PNe (He~1-5, HDW~6 and HaWe~5)  are  removed (see text).}

\end{table*}

\begin{table*}

\caption[]{The results of the $\Sigma-D$ and $L-D$ fits of
PN samples defined using the different methods of distance
determination. The parameters of the fit ($\beta$
and $\alpha$, respectively) and the correlation coefficient $r$
are given for each sample. The number of PNe in the sample is
given in the last column ($N$).}

\smallskip

\centerline{\begin{tabular} {llccccc}
\hline \hline \noalign{\smallskip}
No.& Sample$^{a}$&
$\beta_{_{\Sigma - D}}$  &$r_{_{\Sigma- D}}$&
$\alpha_{_{L-D}}$ & $r_{_{L - D}}$ & $N$ \\
\noalign{\smallskip} \hline \noalign{\smallskip}
1 & Trigonometric Parallax                     & 1.79$\pm$0.22 & $-0.90$ & $-0.21$ $\pm$0.22 &\p00.23  & 18\\
2 & Trigonometric Parallax \dag                & 2.02$\pm$0.18 & $-0.95$ & \p00.02 $\pm$0.18 &$-0.02$  & 17\\
3 & Geometrical (trig. parall. + expan.)       & 2.26$\pm$0.26 & $-0.86$ & \p00.26$\pm$0.26  &$-0.19$  & 30\\
4 & Geometrical (trig. parall. + expan.) \dag  & 2.44$\pm$0.22 & $-0.91$ & \p00.44$\pm$0.22  &$-0.36$  & 29\\
5 & Gravitational                              & 1.42$\pm$0.37 & $-0.66$ & $-0.58$ $\pm$0.37 &\p00.33 & 22\\
\noalign{\smallskip}
6 & Gravitational  \dag                        & 2.53$\pm$0.31 & $-0.89$ & \p00.53 $\pm$0.31 &$-0.38$ & 20\\
7 & Spectroscopic                              & 2.57$\pm$0.36 & $-0.91$ & \p00.57 $\pm$0.36 &$-0.43$  & 13 \\
8 & Extinction                                 & 1.69$\pm$0.38 & $-0.80$ & $-0.31$ $\pm$0.38 &\p00.25 & 13\\
9 & USNO-PN                                    & 2.38$\pm$0.56 & $-0.90$ & \p00.38 $\pm$0.56 & $-0.32$ & \p06\\
\noalign{\smallskip} \hline
\end{tabular}}

\smallskip

$^{a}$ {For the first eight samples, PNe are extracted from the SSV and Ph02; for
USNO-PN -- see Table~3.}

\end{table*}

\begin{table*}
\caption{The data for the USNO PN sample.} \label{tab1e:2}
\centerline{\begin{tabular}{lccc}
\hline \hline \noalign{\smallskip}
Name & Trigonometric Parallax &
 $S_\mathrm{5GHz}$  &
 Diameter  \\
 & [mas] & [mJy] & [$^{\prime\prime}$] \\ \noalign{\smallskip}
\hline
\noalign{\smallskip}
NGC 7293 (036.1-57.1) & $4.56\pm0.49$ & 1292$^a$ & 660$^b$ \\
NGC 6853 (060.8-03.6) & $3.81\pm0.47$ & 1325$^a$ & 340$^a$ \\
NGC 6720 (063.1+13.9) & $1.42\pm0.55$ & \p0384$^a$  & \p060$^c$\\
A 21 (205.1+14.2)     & $1.85\pm0.51$ & \p0157$^d$  & 550$^e$ \\
A 7 (215.5-30.8)      & $1.48\pm0.42$ & \p0305$^a$  & 760$^a$ \\
A 24 (217.1+14.7)     & $1.92\pm0.34$ & \p0\p036$^a$   & 415$^d$ \\
\noalign{\smallskip}
\hline
\noalign{\smallskip}
\end{tabular}}
\small{References: $^a$\cite{Cahn92}; $^b$\cite{1998AJ....116.1346O}; $^c$\cite{1974AA....35..219G};
$^d$\cite{1999AAS..138..275H}; $^e$\cite{1984AA...137..291S}.}
\end{table*}

Malmquist bias tends to increase the slope
of the  $\Sigma-D$ relation for SNRs (\cite{Uro05}).
Similar implications   should be valid for PNe as
well. Slopes from Table~1 suggest that some of our sample sets may
suffer from a Malmquist bias. { Specifically,   $\Sigma-D$ slopes derived
from the SSV and SSV+Ph02 are steeper for higher sampling distances  (i.e. $>2$ kpc).}


\begin{figure*}
\centerline{
\includegraphics[width=16cm]{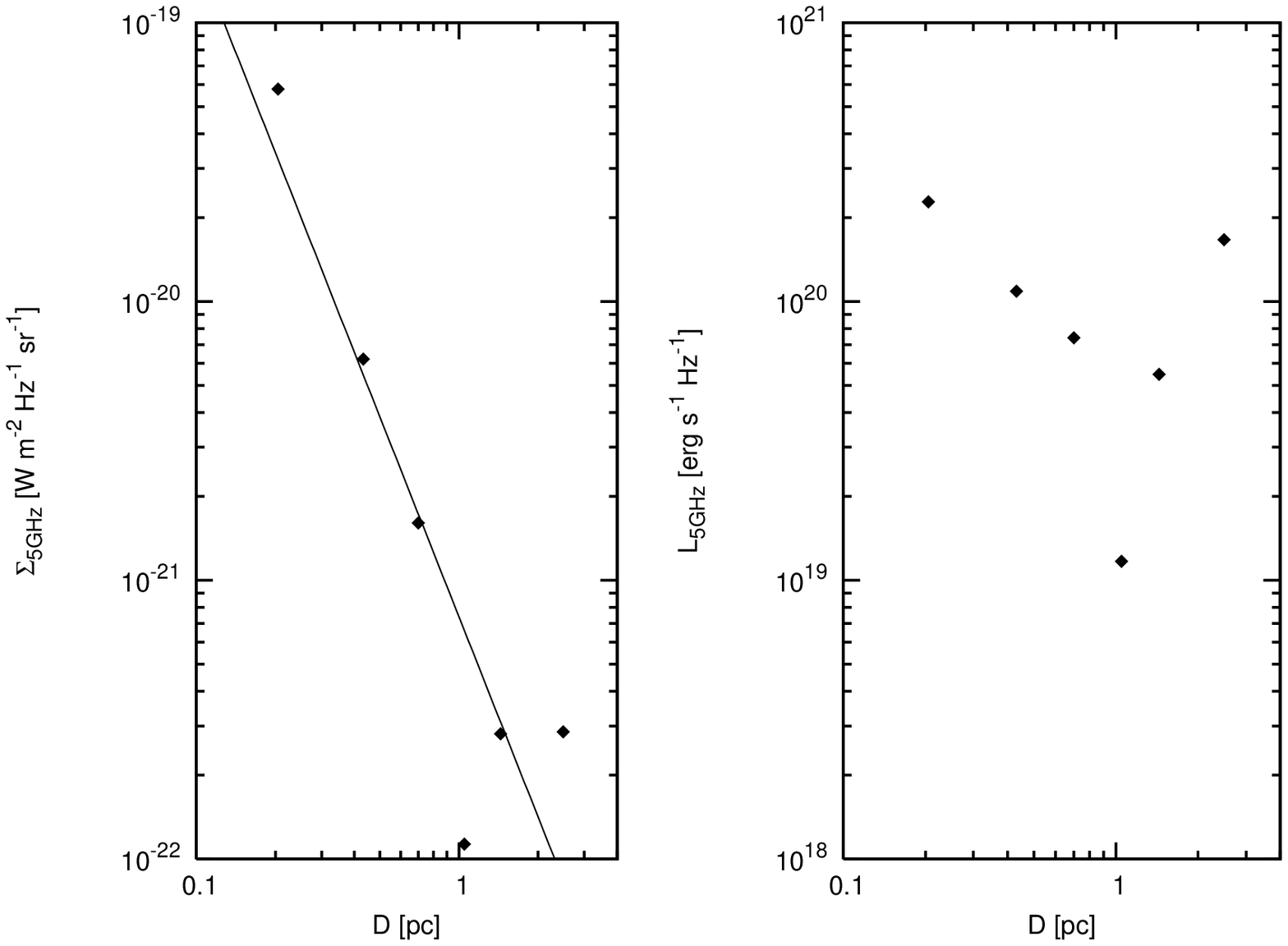}}
\caption{ ({\it Left})   $\Sigma_\nu-D$ diagram at 5 GHz for   sample set 9 in Table 2, having 6
USNO-PNe with accurate trigonometric distances determinations. ({\it Right})
Corresponding 5 GHz $L_\nu-D$ diagram for the same  set of PNe.}
\end{figure*}



In spite of  the possibility of Malmquist
bias, the majority of the  $\Sigma-D$ slopes listed Tables 1 and 2,
are very close to the so-called trivial $\Sigma-D$ relation
with $\beta=2$ (\cite{Arbutina04}).  This can be understood mathematically
from:

\begin{equation}
\Sigma \propto \frac{S}{\Omega}\propto
\frac{L(D)/d^2}{D^2/d^2}\propto D^{-2} L (D),
\end{equation}

\noindent where $S$ is the flux density, $\Omega$ is the solid
angle occupied by the source and $d$ is the source distance. The
usefulness of the $\Sigma-D$ relation can be tested with the correlation
coefficient $r_{\mathrm{L-D}}$ of the $L-D$ relation, which
should approach --1.0 if the data can be explained by a linear relation.

It is clear
from Equation (29) that the $\Sigma-D$ relation has a trivial $\Sigma
\propto D^{-2}$ form in the absence of $L-D$ correlation.
The abundance of low   $r_{\mathrm{L-D}}$ values
as shown in Tables 1 and 2, indicate that the
dependence between $\Sigma$ and $D$ is uncertain. Therefore,
the validity of $\Sigma-D$ relations is diminished even if free
from  Malmquist bias.  { Aside from a few potential sample sets,
including No. 1 listed in Table 1, most should not be used when
calibrating the PN $\Sigma-D$ relation.

Sample No. 1 in Table 1 is the only set with a respectable correlation coefficient,
$r_{\mathrm{L-D}}=-0.67$,  yet
having a low probability of Malmquist bias\footnote{The
correlation coefficients for sets 3 and  9  from Table 1 are the
steepest, but   Malmquist bias is also likely.  All other $\Sigma-D$ slopes are approximately trivial.}.
This SSV  sample consists
of 13 relatively nearby Galactic PNe, with reliable individually
determined distances less than 1 kpc.  They are different in
morphological form  (e.g. round, elliptical, bipolar core,
bipolar, etc.). For this sample, we obtain  the $\Sigma-D$ relation:

\begin{equation}
\Sigma = 7.83^{+4.17}_{-2.73}10^{-22}D^{-2.61\pm0.21},
\end{equation}

\noindent and a corresponding $\Sigma_\nu-D$ and $L_\nu-D$ diagrams
are shown in Fig.~2.

We define the fractional error as:

\begin{equation}
f\!e=\bigg|{d_{\mathrm SSV}-d_{\Sigma}\over d_{\mathrm SSV}}\bigg|,
\end{equation}

\noindent in order to obtain an additional estimate of   accuracy. Independently derived
distances $d_{\mathrm SSV}$ are taken from  SSV and   $d_{\Sigma}$ is the
  distance  derived (for each individual PN) from
the $\Sigma-D$ relation. The maximum and average fractional errors are
$f\!e_{\mathrm max}=1.08$ and $\bar{f\!e}=0.35$, respectively.

\begin{figure*}
\centerline{
\includegraphics[width=16cm]{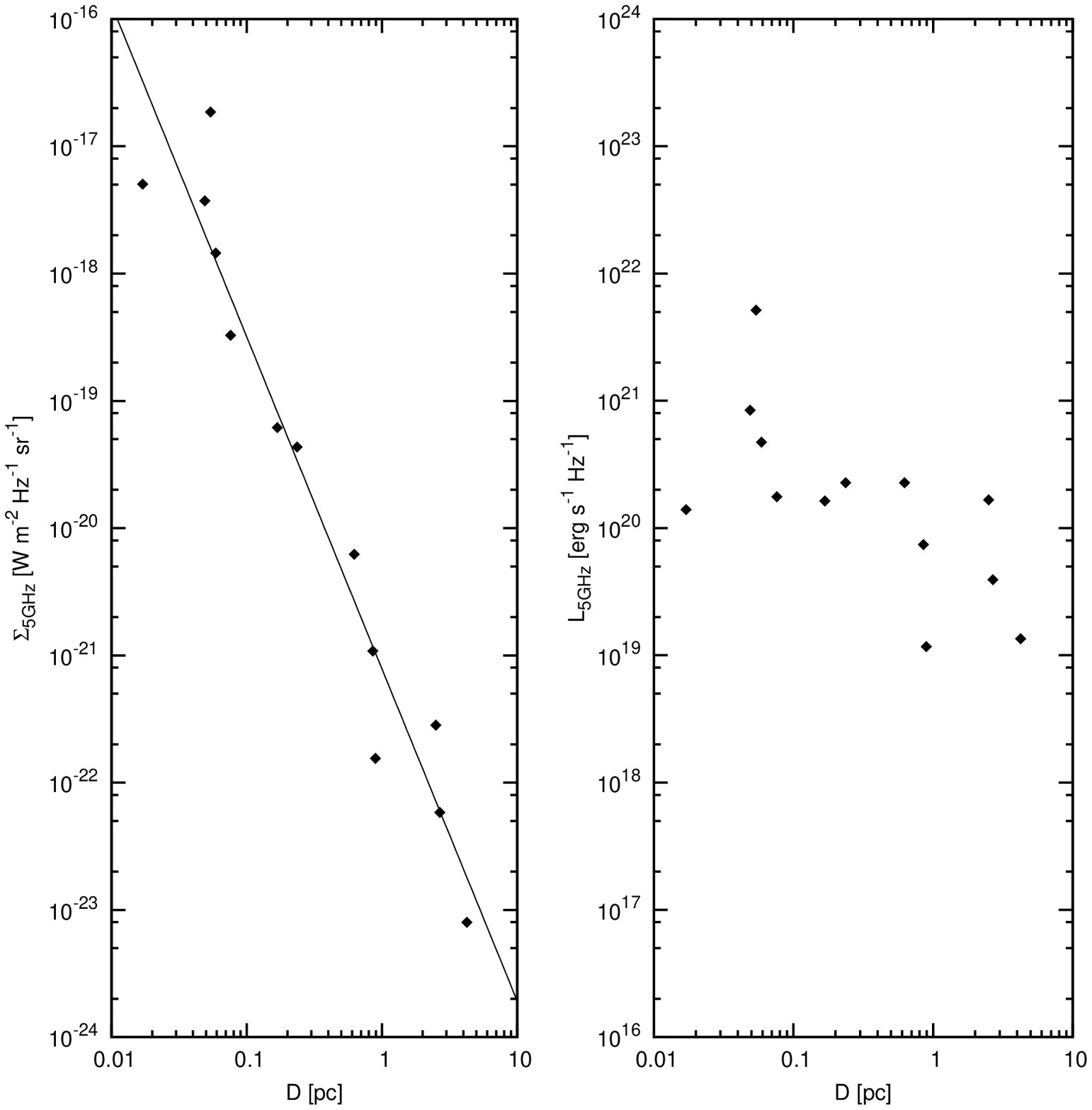}}
\caption{ ({\it Left})  $\Sigma_\nu-D$ diagram at 5 GHz for the set  of 13
SSV PNe with distances $< 1$ kpc (from sample set 1 in Table 1). ({\it Right}) Corresponding
$L_\nu-D$ diagram at~5 GHz for the same set of PNe.}
\end{figure*}

According to our criteria and the relatively small average
fractional error, Equation (30)  may be used as a calibration
relation for distance determination. Since the PNe in the corresponding sample are
nearby, the Malmquist bias should not be significant. The $L-D$
relation exists and the slope $\beta$ is reasonable. However,
this slope is still most likely a "mixture" of the evolutionary and selection
effects. Additionally, the sample number is small (only 13 PNe) and they do
not belong to the same morphological type. Therefore, all results obtained
using Equation (30) should be taken with caution as we discuss below.}

\section{Discussion}


In Section~2, we derived the theoretical dependence between the gas
density in the PN shell and its radius with respect to the
interacting stellar winds model. Next, we showed that the slope of the theoretical $\Sigma-D$
relation is changed significantly by including these equations.

The theoretically derived slope $(\beta\approx1)$ is shallower
than  slopes from the empirical relations given in Tables 1 and
2.  This discrepancy may possibly be explained by   { poor} quality
of   Galactic PN samples or by incorrect assumptions used for
the derivation of the theoretical relation. { Due to variations in the power
law density distribution derived here and the approximately constant
temperature of the expanding PN envelope, the theoretical slope may be
different than given in Equations (26) and (27) (derived
assuming the shell thickness $\Delta\approx$ const.). If the filling factor  $f $ is
constant, the slope $\beta$ could be as great as 3.
  We conclude that   whether these simply derived theoretical relations have the correct forms
or not,   empirical relations are under the influence of biases that make
their slopes different than they would be if one  considers the evolution of   a single hypothetical  PN.
An empirical relation can be useful for estimating distances only under strict conditions as discussed below.}

Van de Steene \& Zijlstra (1995) and Phillips (2004) showed  that the evolution of PNe may  not have a linear trend in
log-log scales. When the radiation of the central star begins to
ionize the nebula, the radio brightness will increase rapidly.
However, in
this phase of PN evolution, the observed incidence of radio nebulae
  is predicted to be low.
    Therefore, our linear trend of
the $\Sigma-D$ form can be justified and the quadratic form of the
$R-T_{\rm b}$ dependence given by \cite{Phillips04} is not necessary
for the statistical analysis given here. In addition, different
dependencies can not be derived from the thermal bremsstrahlung
radiation formula (Equation  24).

{ The subjects of our analysis are optically thin PNe.
The reason for the low incidence of observed radio PN during the rising phase,
discussed above, is that
the majority of these PNe are optically thick at 5~GHz.  Then, when the
central star reaches a temperature of 40 000 K, the number of
ionizing photons becomes almost independent of temperature and
radio luminosity no longer increases.  As discussed by \cite{Van95}, this causes a
decrease in surface brightness with a gradual expansion of the nebula,
which then becomes optically thin and thus detectable.
To keep our sample pure, we devised a set of criteria which
excluded three suspected
thick PNe  in their rising phase of evolution  (as marked with ``\dag" in Table 1).}



{ From   Tables 1 and 2,  we suggest that 24  slopes} are most likely
 trivial $\Sigma-D$ relations. Therefore, we feel that the
Malmquist bias is not so severe as in the case of Galactic SNR
samples, where the slopes are significantly steeper. For example,
the obtained slope for the sample of 132 bulge PNe collected by
\cite{Van95} takes a relatively shallow value of $\beta\approx2$.
This is expected result since all 132 PNe are located at the
approximately same distance.
Malmquist bias does not exist in
samples which contain objects at the same distance.

{ Both \cite{Schneider96} and at a later time, \cite{Bensby01} attempted to improve
the Bulge sample by removing foreground and background PNe. \cite{Bensby01}
defined stronger criteria to extract a more pure Bulge sample (Set 14 in Table 1).
We   found a trivial $\Sigma-D$ slope for
this sample. We  also applied  this stronger criteria directly to the \cite{Van95} sample. The change in slope for this modified Van de
Steene \& Zijlstra sample (Set 18 in Table~1) is negligible. This again supports the idea
that these sample sets are not strongly affected   by the Malmquist bias.}

{ According to   criteria defined in this paper, the corresponding slopes   for
the Bulge samples represent   the result of scattering in the $L-D$ plane.}  This slope
($\beta\approx2$) was obtained for the extragalactic samples of
SNRs (except M82 sample) where Malmquist bias also does not have severe
effect.  Again, this is because all of the SNRs in each set  are at
approximately the same distance (see \cite{Uro02} and  \cite{Uro05}).

{ By examination of sample sets 5 -- 9 in Table 1 (and Fig.~3) we try to demonstrate
the effect of Malmquist bias.
For set 5 with distances less than 0.4~kpc,
the corresponding $\Sigma-D$ slope is $\beta=1.47$.
Further distance comparisons (in the form of distance
interval in kpc followed by the $\Sigma-D$ slope) include:
$0.4<d\leq0.6$, 2.18; $0.6<d\leq1.0$, 2.37;  $1.0<d\leq2.0$, 2.64; and $>2.0$, 4.44.  This implies
that an increase in   distance
gives a $\Sigma-D$ slope that is also greater.

Using both the
Tukey-Kramer (T-K) and the Scheffe (S) methods, we found  the $\Sigma-D$ slope for sample set 9 in Table 1
($>2.0$, 4.44) is statistically different at a confidence level of  $\alpha=0.05$, if
compared to the slopes of sets  5,  6, and  7.
However, since all of these samples
consist of PNe with some uncertainty in their distances,
any firmer conclusion about their Malmquist bias
should be avoided. Using the same
methods, sets 1 ($<1$, 2.61) and  3 ($\geq2$, 3.96)  in Table 1 also represent
a statistically different population at  a confidence level of
$\alpha=0.05$.  The difference between their corresponding  $\Sigma-D$ slopes
is statistically significant at a confidence level of $\alpha=0.2$ and
can be explained by the Malmquist bias.
Sample set 2 in Table~1 is omitted from our slope  analysis
because of its extremely high level of
scattering in the $L-D$ plane (see Table~1, $r_{L-D}=0.08$).


Inspection of Table~2 reveals that while all $\Sigma-D$ slopes are
relatively close to the trivial form, set 4, 6 and 7 are a
bit steeper.   The increased slope of set 4 can
be explained by perturbations induced by twelve very luminous and
highly $\Sigma-D$ plane dispersed   ``expansion'' PNe added to
sample set 2.
Since the   gravitational and spectroscopic
methods are mostly applied for distant PNe, slopes in sets 6 and 7
may be steeper due to a Malmquist bias. However, all of the $L-D$ correlation
coefficients in Table~2  are very low and therefore
poor for   statistical investigations.}

\begin{figure}
{\includegraphics[width=8cm]{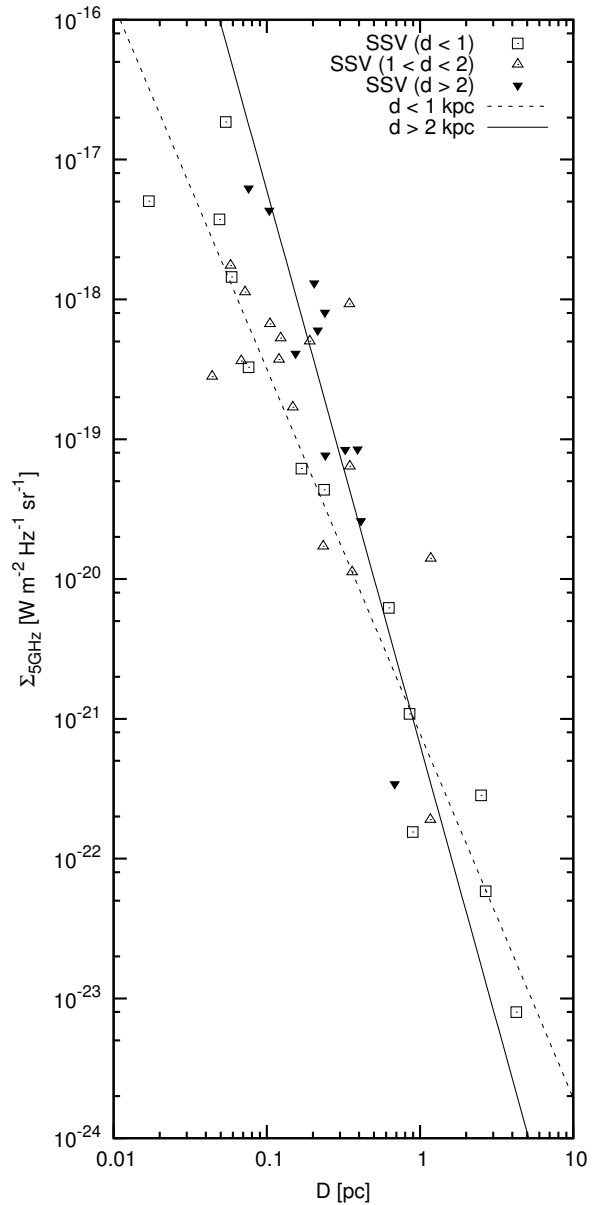}}
\caption{The $\Sigma_\nu-D$ diagram at 5 GHz for all 39 SSV PNe
(sample set 4) with   reliable individually determined distances. The
straight lines represent the least square fit lines through the
two distance limiting Galactic samples. The difference
in slope  of corresponding $\Sigma-D$ relations are
defined for PNe with distances $<1$ kpc and $\ge2$ kpc (represented
with dash and tick line, respectively).  These could be explained as a result of
 the  volume selection effect called Malmquist bias.}
\end{figure}

The   high level of scattering in the $L-D$ plane, as seen in Figs.~1
(right) and 4,  supports the idea that most of the slopes in Tables~1
and 2 do not have any real physical interpretation. It is this
  luminosity-diameter scattering artifact which produces the
trivial $\Sigma\propto D^{-2}$ form. Therefore,   { 24} of these trivial
$\Sigma-D$ slopes are not useful
for the determination of   distances. Causes for this
scattering include: imprecisely   determined calibrator distances,
mixtures of different  PN types in the same
sample,   limitations in sensitivity and resolution of radio
surveys,   source confusion  and to a lesser extent,
Malmquist bias.

{ Although, our criteria imply that the
relation given in Equation (30) could be useful for distance determination, we assign a low confidence
that it represents a real evolutionary track.
  The $\Sigma-D$ slope from this relation represents the result of coupling of evolutionary
  and selection effects. The corresponding sample set consist of only 13 PNe
  and they do not belong to the same
morphological type. Thus, this sample is not complete or homogeneous and
distance estimates obtained by using Equation (30) should be viewed with appropriate  caution.

PN radio surface brightness depends on a number of factors that include:   intrinsic nebula evolution,  changes in density  and temperature, changes in medium ionization  and    extremely large luminosity variations in   the central star during its
evolution. In Section~2, we assumed that
shell expansion was constant  throughout each phase of nebula evolution. On the contrary,
PN shells are accelerated by the central star fast winds at the beginning of PN evolution, and only  later
 are they driven at approximately constant speed  by the hot bubbles created by the
crossing of the fast winds through the inner shocks. \cite{Shaw01}  discusses acceleration of PN shells, i.e. how the velocity of PN shells is
related to the evolution of the central star. They emphasized that the acceleration of   shells is nearly
zero in larger (and probably older) PNe having diameters $> 0.03$ pc. Almost all of the PNe from
 sample sets presented in Tables 1 and 2  are larger than 0.03 pc (see Fig. 4)  and therefore  in
later phases of evolution where acceleration (or central star influence) is negligible.
This explanation is admittedly simplistic. For essential understanding of the theoretical and empirical $\Sigma-D$ dependencies,
the full explanation of the complex co-evolution of a nebula and its central star would be needed.

The most massive central stars will never fully ionize their environment and those PNe will never become optically
thin at 5~GHz since their central stars evolve  so rapidly. Those observed PNe with   smaller
diameters and surface brightnesses (optically thick PNe at 5~GHz) are located significantly lower in the
$\Sigma-D$ or $L-D$ planes when compared to the average (thin) PNe.  Their evolution are surely not defined by the
$\Sigma-D$ relations derived here. Although the majority of the thick PNe will become thin,   the question
is WHEN?  This depends on a number of factors during their evolution.
These    effects   significantly contribute to   uncertainties in addition to the huge scattering found in the $L-D$ plane.

Other factors not included in our analysis
may have a large impact on the evolution of PN emissivity. These   include: compression of   gas induced by
the interaction of   fast wind with the medium on the main shell inner boundary,   isothermal shocks associated
with the passage of   D-ionization fronts through neutral gas  and progressive ionization of greater amounts of
surrounding neutral gas (e.g. \cite{Shaw06}, and references therein). We do not analyze   ionization fronts in this paper; our derivations in Section~2 are based only on the dynamics of corresponding shock waves.
The surface brightness evolution may be more closely connected to ionization than the
dynamical expansion of the shell. This may be good enough  reason to challenge the quality of our derived
theoretical relations and an additional explanation for scattering in the $L-D$ plane.
Finally, we concede that some
PNe could have different initial conditions   leading to
independent evolutionary paths. These paths could follow the same
theoretical $\Sigma-D$ curve but with varying intercepts,
therefore leading to the observed scatter.}

In order to define a valid PN sample for
statistical investigation, one needs to separate a higher number
of nearby PNe with the same morphological characteristics.
We define  such a sample (USNO-PN) in the
subsection below (Section~5.1). However, the quality of radio-continuum
observations today  simply is not sensitive enough to provide  a high  number of similar PNe  { having reliably
determined distances}.

\subsection{The USNO-PN Sample}

We have examined the morphological properties of PNe from the USNO-PN  sample
using the radio images of these objects found in the literature. We use comparisons
with   radio morphological classes given by \cite{1996ApJ...462..813A} which are
based on the {\it prolate ellipsoidal model} (PES). The PES model describes divergency
in observed (empirical) morphologies through   different sky projections of
relatively simple spherical shell models with both radial and latitude density
gradients and with different ionization depths.  USNO-PN  sample objects can
be classified as follows: circular  (NGC 7293 or Helix Nebula, see \cite{1989AAS...79..329Z});
open elliptical (NGC 6853 or Dumbbell Nebula, see \cite{1983IAUS..103...69B});
elliptical  (NGC 6720 or Ring Nebula, see \cite{1974AA....35..219Gnp});
symmetric (A21 or S274, see \cite{1984AA...137..291Snp}); and S-type (A24\footnote{For   A24 we acquire a 1.4~GHz radio image from the
NRAO VLA Sky Survey (NVSS) postage stamp server (http://www.cv.nrao.edu/nvss/postage.shtml).}).
However, a further search of the literature reveals that the intrinsic
morphologies for all five of these objects is most likely to be   bipolar, i.e.
that the structures of the emitting regions (or shells) range from a thick disk
(Helix Nebula) to a triaxial ellipsoid (A24, and Ring Nebula) to
a barrel-like shape (A21, and Dumbbell Nebula) (see \cite{1998AJ....116.1346Onp,
1999ApJ...517..782H,1999AAS..138..275Hnp,2007AJ....134.1679O} and \cite{2005RMxAA..41..109M}).
The sixth object for which we found reliable radio observations (A7) appears to be
a ``classical'' PN with a well defined spherical shell (\cite{1996AA...310..603X}).

We   note that \cite{1996ApJ...462..777K} and \cite{1988ApJ...324..501Z}
detected a shock-excited H$_2$ emission from the Helix, Ring and Dumbbell
nebulae, implying an ionization bounded case for these objects. Additionally, no
spectroscopic evidence for the existence of a high-velocity stellar wind has been
found for the previously mentioned objects, including A7
(\cite{1985ApJ...291..237C,1991AAS...91..325P}).
Central stars of these objects are most likely  well pass the
hydrogen-shell burning phase and approaching the white dwarf cooling sequence.

A wide range of shell structures is predicted in the hydrodynamical models
based on the ISW model and it is very likely that a large number of
PNe classified by  apparent morphology as   spherical or   elliptical,
may in fact posses a bipolar structure. Even though morphological
classifications based on   optical imaging favor elliptical objects,
\cite{1986ApJ...301..772Z} predict that approximately
50\% of all PN in the Galaxy are actually bipolar. Nevertheless, we conclude
that the USNO-PN sample is not representative of the majority of PNe because
of its ``narrowness'' in morphology (i.e. almost all  appear to be
bipolar).

\section{Summary}

The main results of this paper may be summarized as follows:

\begin{enumerate} [(i)]

\item We have derived a new theoretical $\Sigma-D$
relation for PNe in the form $\Sigma\propto D^{-1}$  by including
the interaction between the AGB star wind and the fast wind from the
central star. This dependence is obtained for both  momentum
conserving and   energy conserving phases of evolution. Only
for the final stage, we did derive the same theoretical dependence
($\Sigma\propto D^{-3}$) as   in Paper I.


\item We derived   empirical $\Sigma-D$ relations for   { 6 updated PN
samples  from the literature, 12 new sample sets   extracted from these and 9
additional sets as defined in this paper}. We discuss the selection effects that influence these
PN sample sets. Our results show that   updated Galactic PN
samples   do not severely suffer from volume selection effects, i.e.
Malmquist bias (similar as in case of the extragalactic SNR samples).
{ We do note that the $\Sigma-D$ slope does increase with sample distance,
possibly due to Malmquist bias as shown on two statistically different  SSV  samples with distance
limits $<1$ and $\geq2$ kpc.
Because of this trend, we feel that samples consisting only of nearby PNe are useful for
future statistical analysis.}

\item  { From the analysis of $L-D$
dependencies presented here, we conclude that   24 empirical
$\Sigma-D$ relations (listed in Tables 1 and 2) for Galactic PNe should not
be used to determine   distances to other  observed PNe
  or to establish  any other PN evolutionary paths. These
relations represent only a general trend of changes in surface
brightness. Two obtained steep $\Sigma-D$ slopes for distant samples
($d>2$ kpc) are probably the result of  Malmquist bias.
However, we extracted one useful nearby  SSV subsample ($d<1$ kpc, set 1
in Table 1). This  updated sample   consists
of PNe with reliable and individually determined distances. A reasonable $L-D$
dependence does exist  for this nearby  sample. Since this  sample
consists of only 13 morphologically different PNe, the
corresponding $\Sigma-D$ relation for  distance  determination
must be used  with caution.}

\end{enumerate}


\begin{figure*}
\centerline{
\includegraphics[width=16cm]{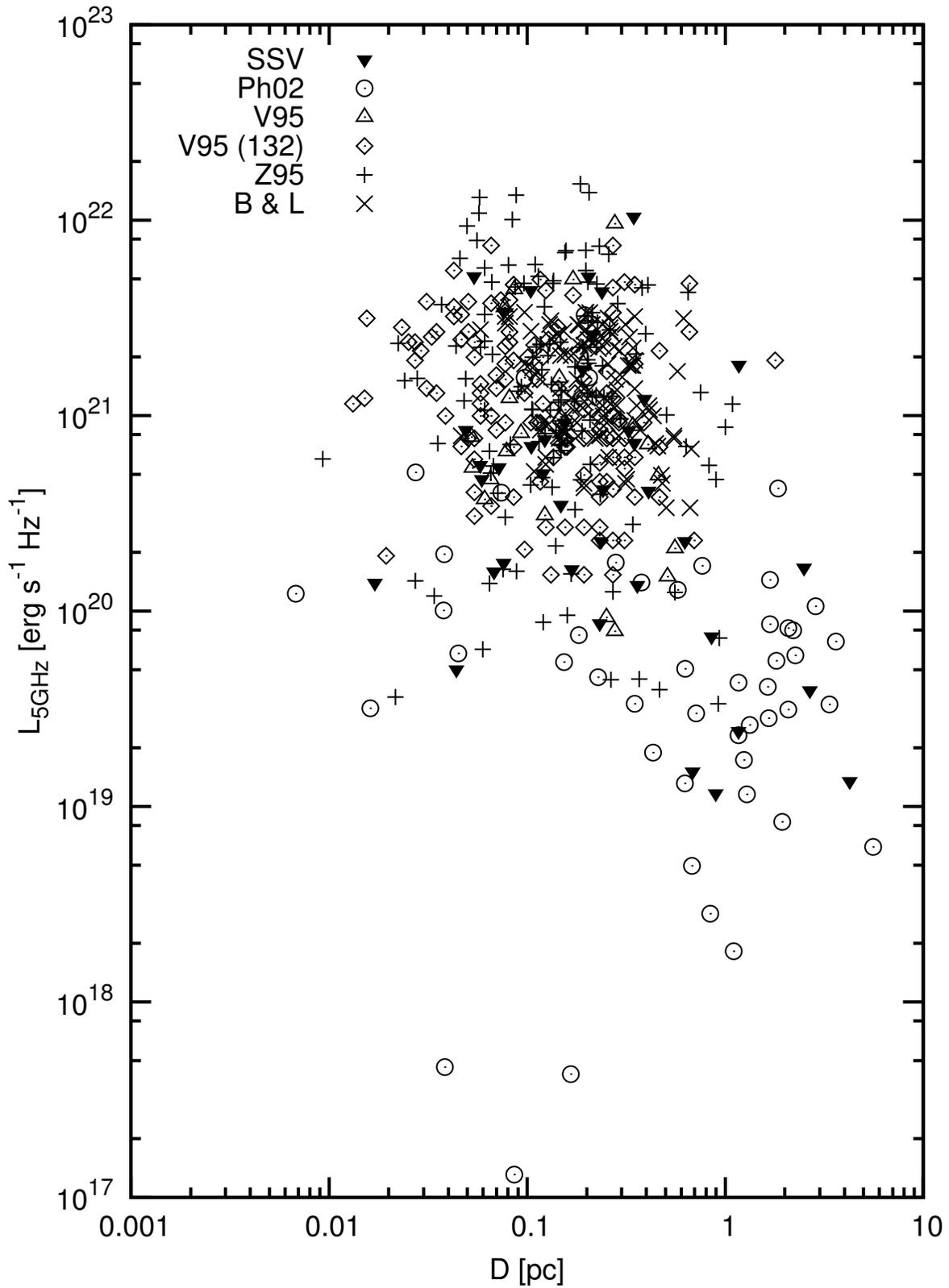}}
\caption{The $L_\nu-D$ diagram for the samples of PNe given in
Table 1.}
\end{figure*}

\begin{acknowledgements}

 { We thank an anonymous referee for many helpful
comments that substantially improved the quality of this paper. The authors thank Jeffrey Payne for careful
reading and correction of the manuscript. This research has made
use of the SIMBAD database operated at CDS in Strasbourg, France,  NASA's Astrophysics Data System and
  has been supported by the Ministry
of Science and Environmental Protection of the Republic of Serbia
(Projects: Nos. 146002,   146003,  and 146012).}

\end{acknowledgements}

\end{document}